\documentclass[showpacs,preprintnumbers,amsmath,amssymb,twocolumn,prb]{revtex4}

\ifx\pdftexversion\undefined
  \usepackage[dvips]{graphics}
\else
  \usepackage[pdftex]{graphics}
\fi

\usepackage{graphicx}

\begin{document}

\title{Photovoltaic effect in bent quantum wires}

\author{Yuriy V. Pershin}
\email{pershin@pa.msu.edu}
\author{Carlo Piermarocchi}
\affiliation{Department of Physics and Astronomy, Michigan State
University, East Lansing, Michigan 48824-2320, USA}

\begin{abstract}
We propose a scheme for the generation of photocurrent in bent quantum
wires. We calculate the current using a generalized
Landauer-B\"uttiker approach that takes into account the
electromagnetic radiation. For circularly polarized light, we find
that the curvature in the bent wire induces an asymmetry in the
scattering coefficients for left and right moving electrons. This
asymmetry results in a current at zero bias voltage.  The effect is
due to the geometry of the wire which transforms the photon angular
momentum into translational motion for the electrons.
\end{abstract}

\pacs{73.21.Hb, 72.40.+w, 72.30.+q} \maketitle

 Recently many schemes for photocurrent generation in confined
electron systems have been investigated.~\cite{lit2}$^-$\cite{lit}
In particular, mechanisms of photocurrent generation by circularly
polarized radiation have been considered in quantum
rings~\cite{magarill,perpier2} and helical quantum
wires.~\cite{magarill1} These geometries are particularly
interesting because they can transform angular momentum (from the
photon circular polarization) to translational motion (electron
current).~\cite{magarill} The experimental realization of these
schemes has not yet been reported. In fact, the detection of the
photocurrent in isolated quantum rings is experimentally
challenging and helical quantum wires cannot be fabricated using
standard growth techniques.

\begin{figure}[b]
\centering
\includegraphics[angle=270,width=8.5cm]{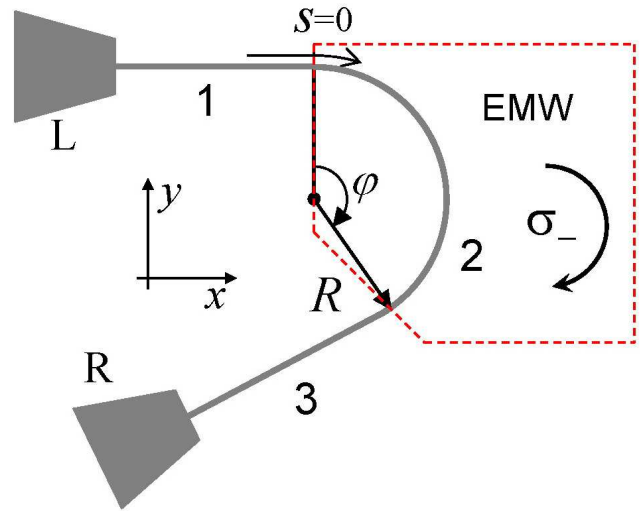}
\caption{(Color online) Curved quantum wire irradiated by a
circularly polarized electromagnetic wave (EMW) with the electric
field component precessing in $(x,y)$-plane.} \label{fig1}
\end{figure}

In this paper, we consider the photocurrent induced in ballistic
quantum wires bent as in Fig.~\ref{fig1}. Quantum wires of this
geometry can be easily fabricated using standard semiconductor
growth techniques, like for instance V-grooving.~\cite{kapon} The
setup in Fig.~\ref{fig1} can also be realized by bending a single
carbon nanotube on a surface. Using a scattering theory approach
we show that the photocurrent can be strong in this geometry. In a
GaAs based quantum wire under a radiation of 33mW/cm$^2$ we obtain
a current of the order of 10 pA, which is measurable with standard
methods. The circularly polarized radiation propagating
perpendicularly to the wire plane induces on the electrons in the
curved region a sliding potential of the form $V(s/R \pm \omega
t)$, where $s$ is the position in the wire, $R$ is the radius of
curvature, and $\hbar \omega$ is the radiation energy.  This
sliding potential is an asymmetric scattering potential for left
and right moving electrons, and the difference in the transmission
probabilities results in a steady current. The classical
interpretation of the effect is that only the electrons moving in
the same direction of the sliding potential are accelerated. We
found that quantum interference plays an important role in the
current. In fact, the energy dependence of the current shows not
only a peak at the Fermi energy, expected from a classical
picture, but also several additional peaks. These additional peaks
are due to the quantum interference of transmitted and reflected
waves at the points where the curvature of the wire changes.

As shown in Fig. \ref{fig1}, we model the curved quantum wire in
the ($x,y$) plane using two straight quantum wires (regions 1 and
3) connected by an arc of a radius $R$ (region 2). On the opposite
side the straight quantum wires are connected to the left (L) and
right (R) electron reservoirs. The arc length is given by
$L=\varphi R$, where $\varphi$ is the arc angle. We assume that a
circularly polarized electromagnetic radiation propagates in the
$z$ direction, perpendicular to ($x,y$) plane. Experimentally, an
electromagnetic cavity can be used to confine the radiation only
in the bent segment of the wire. The electromagnetic cavity will
also enhance the radiation-electron coupling and increase the
current. In any case, a radiation acting on the straight segments
of the wire will not generate a current alone, no matter what
polarization is used. We will therefore neglect the effect of the
radiation on the straight segments. The electron motion along the
curved wire is one-dimensional and we define a parameter $s$ which
indicates the position along the wire. For simplicity, let us
select the point $s=0$ at the contact of regions 1 and 2. The
single electron Hamiltonian in the effective-mass approximation is
given by
\begin{equation}
H=-\frac{\hbar^2}{2m^*}\frac{\partial ^2}{\partial
s^2}+\left(\theta(s)-\theta(s-L)\right)\left[-\boldsymbol{dE}-U_g
\right], \label{ham}
\end{equation}
where $m^*$ is the effective mass, $\theta (..)$ is the step
function, and $\boldsymbol{d}=-e\boldsymbol{r}$ is the dipole
moment. The electric field in the radiation is written as
$\boldsymbol{E}=E_0\cos (\omega t)\hat{x}\pm E_0\sin (\omega
t)\hat{y}$, where $E_0$ and $\omega$ are the electric field
amplitude and frequency, $\hat{x}$ and $\hat{y}$ are unit vectors
in the $x$ and $y$ directions, and $\pm$ corresponds to a
$\sigma_\pm$ circular polarization. The first term in
Eq.~(\ref{ham}) is the kinetic energy, the second term is the
dipolar interaction with the radiation and the third term is the
geometrical potential $U_g=\hbar^2/(8m^*R^2)$ which describes the
effect of the curvature.~\cite{costa81,shevchenko01} The factor
$(\theta(s)-\theta(s-L))$ makes the second and third terms
different than zero only in the bent segment (region 2). Moreover,
we assume that the curved quantum wire is narrow in the transverse
directions, so that our model is limited only to transitions
within the lowest transverse subband.

Using the substitution $y=R\cos(s/R)$ and  $x=R\sin(s/R)$, we can rewrite
the interaction term in Eq. (\ref{ham}) as
\begin{equation}
-\boldsymbol{dE}=2eRE_0\sin\left( \frac{s}{R}\pm\omega t\right)
\label{intpot}.
\end{equation}
According to Eq. (\ref{intpot}), the electrons in the constant
curvature segment are subjected to a potential that moves forward or
backward depending on the helicity of the circularly polarized light.
A similar sliding potential describes the interaction of electrons
confined by a traveling acoustic wave.~\cite{galperin} In our case,
the effective wavelength of the confined travelling wave is $2 \pi
R^{-1}$.

We can write the electric current from the left (L) to the right (R)
reservoirs using a generalization \cite{datta,lit5,galperin} of the
Landauer-B\"uttiker formula \cite{landauer} that takes into account
the radiation
\begin{eqnarray}
I=\frac{2e}{h}\sum\limits_n \int\limits_0^\infty \left[
T_{R,L}(E+n\hbar\omega,E)f_{\mu_L}- \right. \nonumber \\ \left.
T_{L,R}(E+n\hbar\omega,E)f_{\mu_R} \right] dE. \label{curr}
\end{eqnarray}
Here $e$ is the electron charge and $T_{R,L}(E+n\hbar\omega,E)$ is the
probability that an electron of energy $E$ in the left reservoir is
transmitted to the right reservoir in a state of energy
$E+n\hbar\omega$.  Since we are going to study the current in the
absence of external bias, i.e., at $\mu_L=\mu_R=\mu$, Eq. (\ref{curr})
can be rewritten as
\begin{equation}
I=\frac{2e}{h}\int\limits_0^\infty  \Delta T(E)f_{\mu}dE,
\label{curr1}
\end{equation}
where
\begin{equation}
\Delta T(E)=\sum\limits_n\left[ T_{R,L}(E+n\hbar\omega,E)-
T_{L,R}(E+n\hbar\omega,E) \right]. \label{te}
\end{equation}

We first consider the time-dependent Schr\"odinger equation in region
2. Taking into account only single photon absorption and emission
processes, corresponding to $n=-1,0,1$ in Eq.~(\ref{te}), we write the
electronic wave function in the form
\begin{equation}
\psi_2(s,t)=\sum\limits_{n=-1}^{1}f_n(s)e^{-\frac{i(E+n\hbar\omega)t}{\hbar}}.
\label{wavef}
\end{equation}
In what follows we consider the case of $\sigma_-$ polarization, as
shown in Fig. \ref{fig1}. The $\sigma_+$ case is
analogous. Substituting (\ref{wavef}) into the time-dependent
Schr\"odinger equation and neglecting the terms related to
multi-photon absorption and emission, we obtain
\begin{eqnarray}
(E-\hbar\omega+U_g)f_{-1}+\frac{\hbar^2}{2m^*}f_{-1}''=ieE_0Re^{-i\frac{s}{R}}f_0,
\label{eqf1}\\
(E+U_g)f_0+\frac{\hbar^2}{2m^*}f_0''=ieE_0R\left(e^{-i\frac{s}{R}}f_1-e^{i\frac{s}{R}}f_{-1}\right),
\label{eqf2}
\\
(E+\hbar\omega+U_g)f_1+\frac{\hbar^2}{2m^*}f_1''=-ieE_0Re^{i\frac{s}{R}}f_0.
\label{eqf3}
\end{eqnarray}
By looking for solutions of the Eqs. (\ref{eqf1}-\ref{eqf3}) in the
form $f_{-1}=C_{-1}e^{i\left(\tilde{k}-\frac{1}{R}\right)s}$,
$f_0=C_0e^{i\tilde{k}s}$, and
$f_1=C_1e^{i\left(\tilde{k}+\frac{1}{R}\right)s}$, we obtain for the
coefficients the system of equations
\begin{eqnarray}
(E-\hbar\omega+U_g)C_{-1}-\frac{\hbar^2\left(\tilde{k}-\frac{1}{R}\right)^2}{2m^*}C_{-1}-
\nonumber \\ ieE_0RC_0=0, \label{coef1}\\
(E+U_g)C_0-\frac{\hbar^2\tilde{k}^2}{2m^*}C_0+ieE_0R\left(C_{-1}-C_1\right)=0,
\label{coef2}
\\
(E+\hbar\omega+U_g)C_1+\frac{\hbar^2\left(\tilde{k}+\frac{1}{R}\right)^2}{2m^*}C_1+
\nonumber \\ ieE_0RC_0=0. \label{coef3}
\end{eqnarray}
From the condition that the matrix in the linear system of
Eqs.~(\ref{coef1}-\ref{coef3}) has the determinant equal to zero,
we obtain the six possible values of $\tilde{k}$. Therefore, the
wave function in the second region is given by
\begin{equation}
\psi_{2}(s,t)=\sum\limits_{j=1}^6\sum\limits_{n=-1}^1C_{n,j}e^{i\left(\tilde{k}_j+n\frac{1}{R}\right)s}e^{-\frac{i(E+n\hbar\omega)
t}{\hbar}}. \label{wavef2}
\end{equation}
where the coefficients $C_{-1,j}$ and $C_{1,j}$ can be expressed
through $C_{0,j}$ using Eqs. (\ref{coef1}) and (\ref{coef3}).

\begin{figure}[t]
\centering
\includegraphics[angle=270,width=8.5cm]{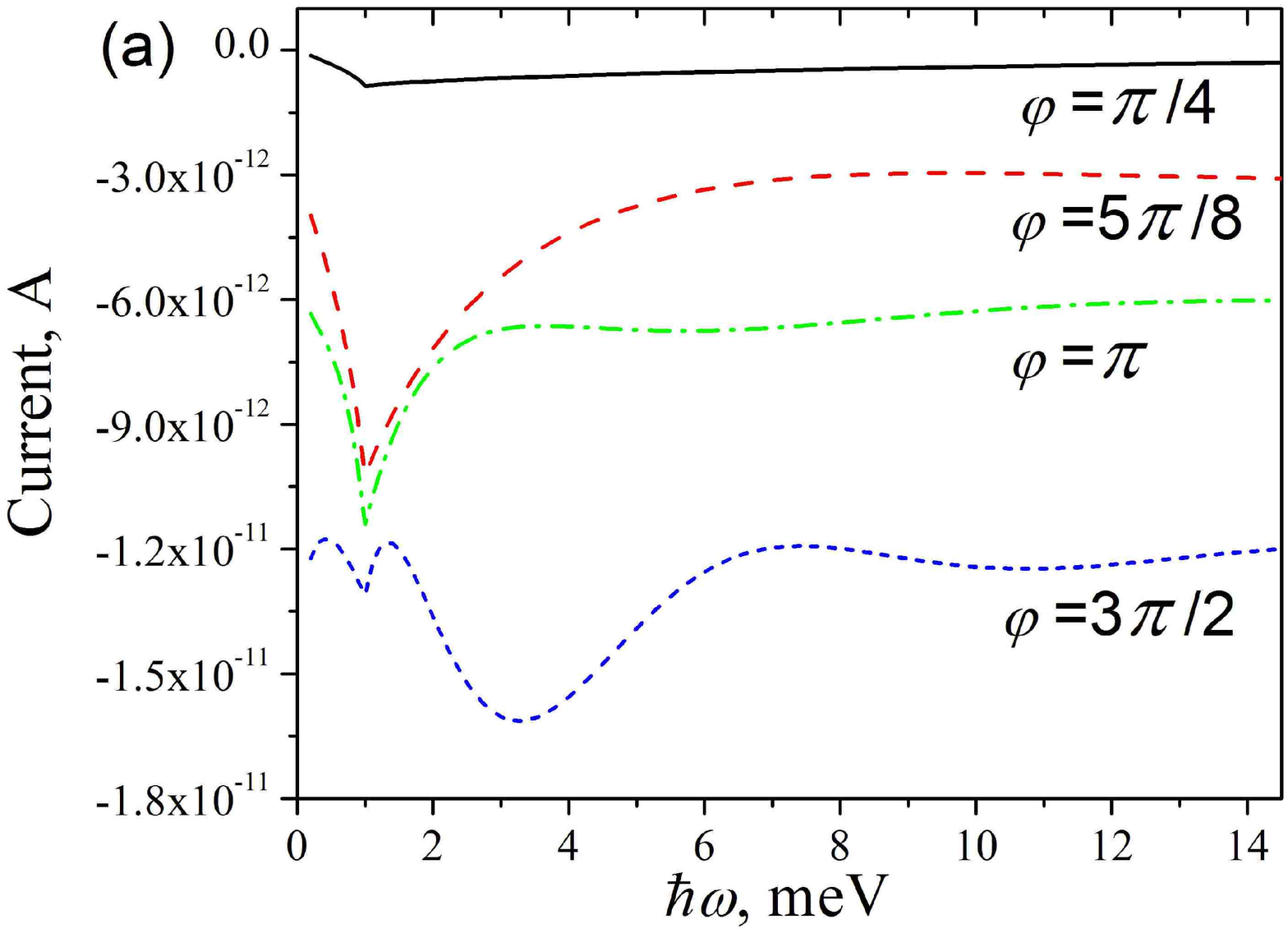}
\includegraphics[angle=270,width=8.5cm]{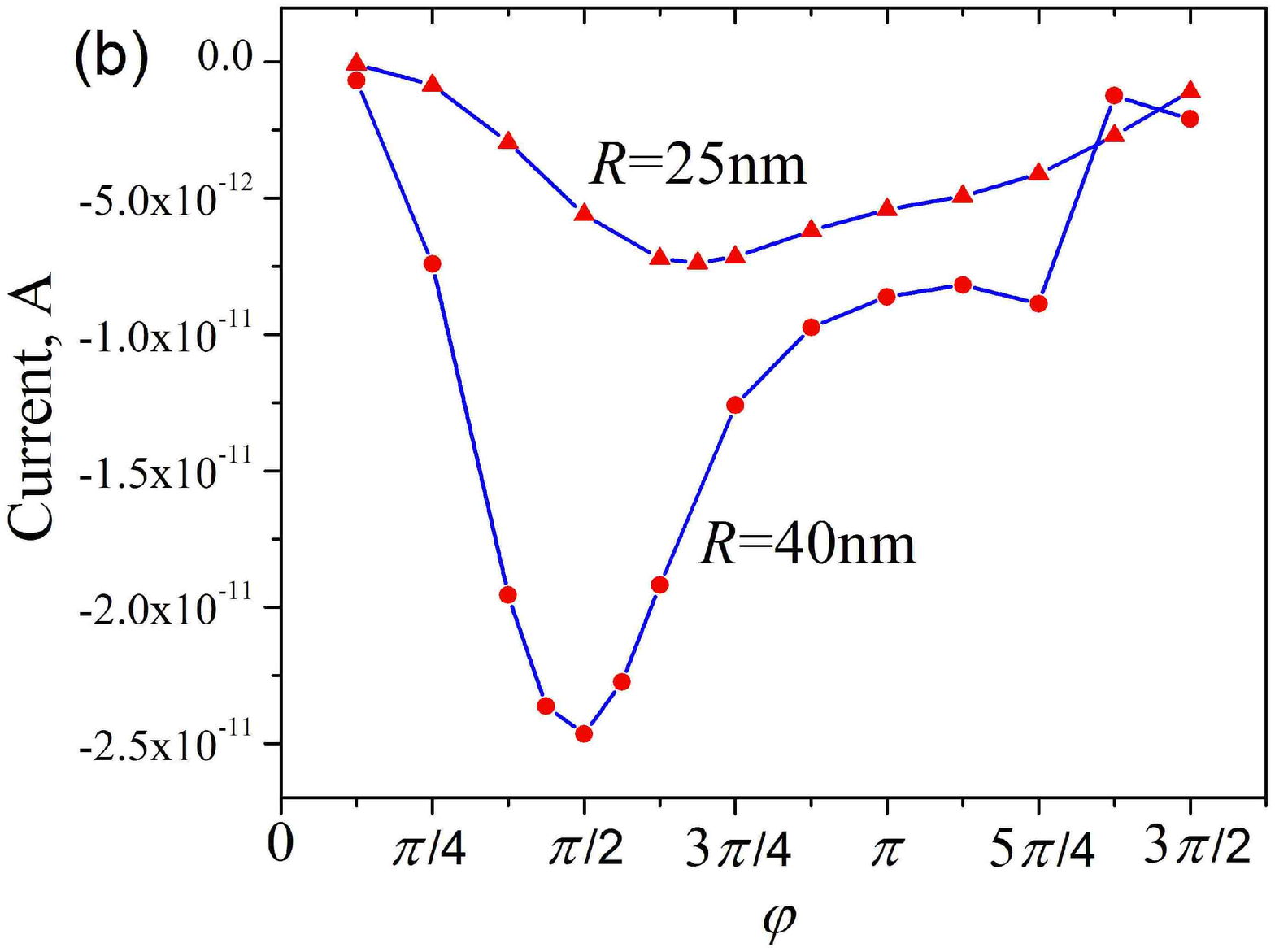}
\caption{(Color online) (a) Photocurrent as a function of the
photon energy calculated for quantum wires having different arc
lengths and $R=25$nm. (b) Photocurrent as a function of the arc
length at $\hbar\omega=1$meV. These plots have been obtained using
$E_0=500$V/m, $\mu=1$meV, $T=10$mK, and $m^*=0.067m_e$. The
electric field amplitude $E_0=500$V/m corresponds to 33mW/cm$^2$
radiation power. All the curves other than $\varphi=\pi /4$ in
Fig. 2 (a) have been vertically shifted by steps of $0.3 \cdot
10^{-11}$A. } \label{fig2}
\end{figure}
\begin{figure}[t]
\centering
\includegraphics[angle=270,width=8.5cm]{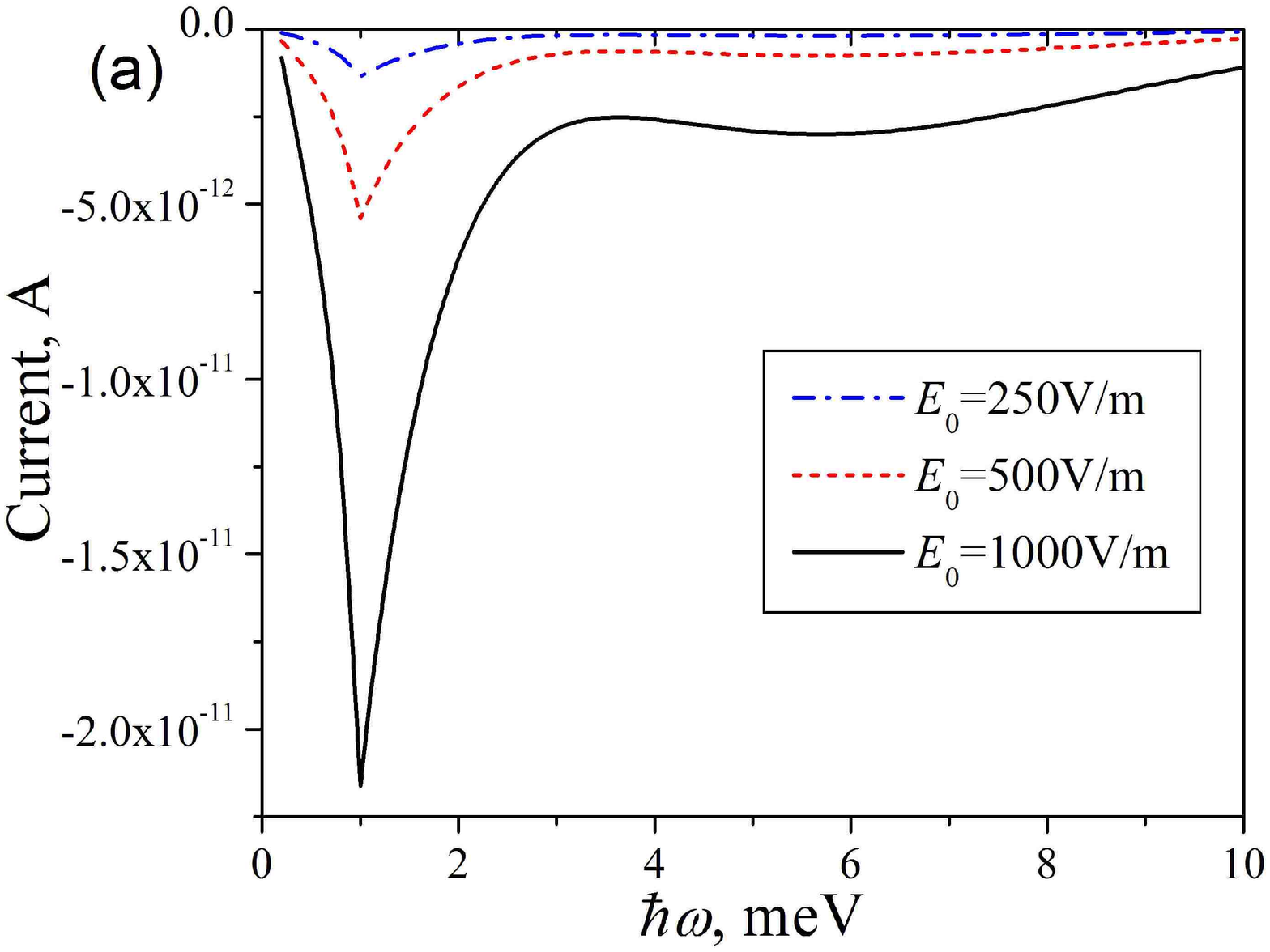}
\includegraphics[angle=270,width=8.5cm]{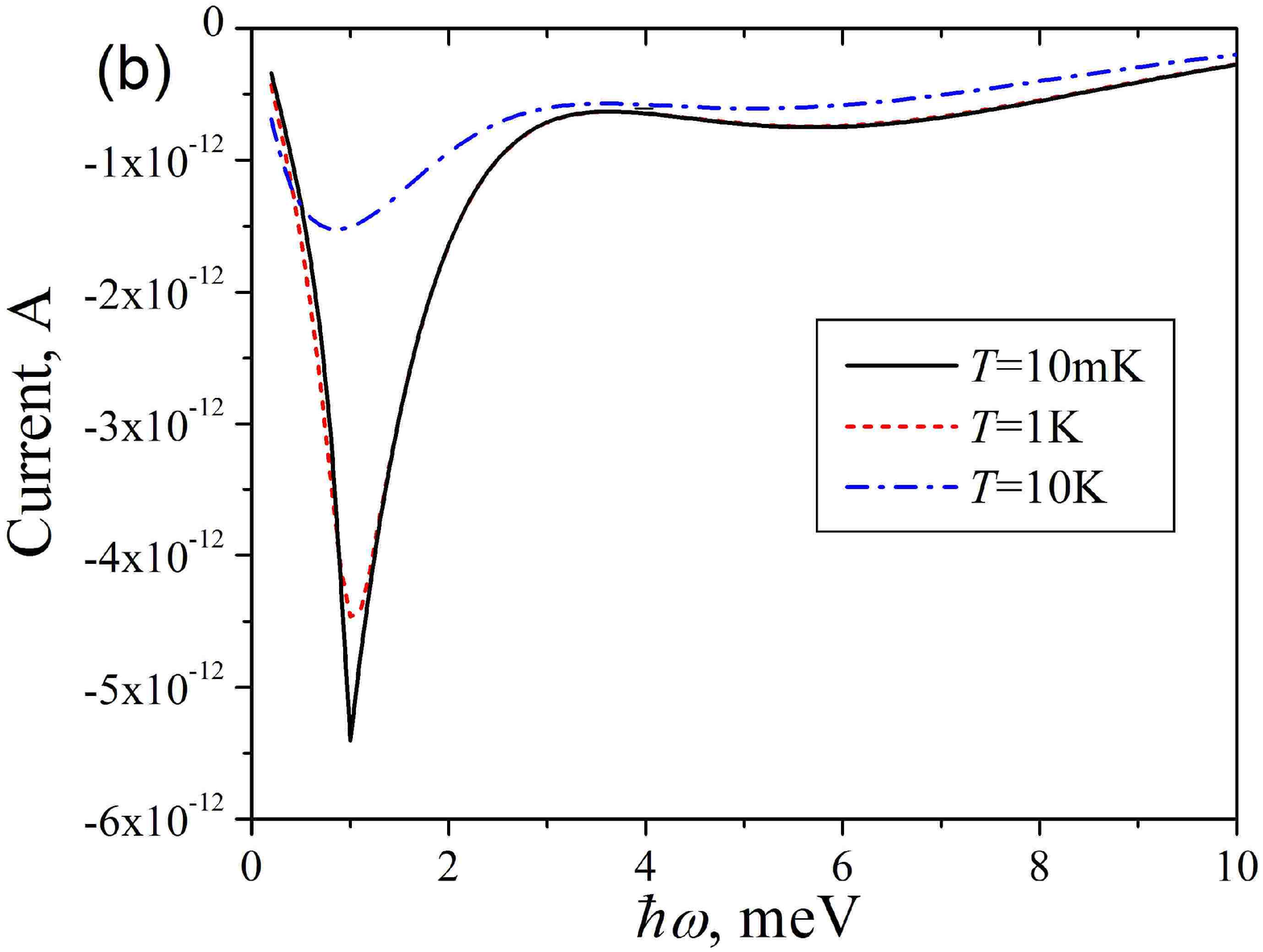}
\caption{(Color online) Photocurrent as a function of the photon
energy calculated (a) for different radiation intensities at
$T=10$mK and (b) for different temperatures at $E_0=500$V/m. The
other parameters values are as in Fig. \ref{fig2} with
$\varphi=\pi$ and $R=25$nm.} \label{fig3}
\end{figure}

In order to find the transmission probability $T_{L,R}$, we solve a
scattering problem selecting the wave functions in the 1-st and 3-rd
regions as
\begin{equation}
\psi_{1}(s,t)=e^{ik_0s}e^{-\frac{iEt}{\hbar}}+\sum\limits_{n=-1}^1r_ne^{-ik_ns}e^{-\frac{i(E+n\hbar\omega)
t}{\hbar}}, \label{wavef1}
\end{equation}
and
\begin{equation}
\psi_{3}(s,t)=\sum\limits_{n=-1}^1t_ne^{ik_ns}e^{-\frac{i(E+n\hbar\omega)
t}{\hbar}}, \label{wavef3}
\end{equation}
where $k_n=\sqrt{2m^*\left(E+n\hbar\omega\right)/\hbar^2}$, $r_n$ and
$t_n$ are reflection and transmission coefficients,
respectively. Matching the wave functions and their derivatives at the
boundaries $s=0$ and $s=L$, we obtain 12 linear equations for the
coefficients $r_n$, $t_n$ and $C_{0,j}$ with $n=-1,0,1$ and
$j=1,..,6$. These equations were solved numerically, and the total
transmission coefficient from the left to the right reservoir at the
energy $E$ was calculated as
\begin{eqnarray}
\sum\limits_{n=-1}^1
T_{R,L}(E+n\hbar\omega,E)=|t_0|^2+\frac{k_1}{k_0}|t_1|^2+ \nonumber \\
\theta\left( E-\hbar\omega\right)\frac{k_{-1}}{k_0}|t_{-1}|^2.
\end{eqnarray}
Using a similar scheme, we obtained the total transmission
coefficient in the opposite direction and calculated the current
using Eq. (\ref{curr1}).

The results of our calculations are shown in Fig.~\ref{fig2} and
Fig.~\ref{fig3}. Typically, the photocurrent is negative, in agreement
with the picture that the potential sliding to the right (for
$\sigma_-$ polarization) increases the transmission probability for
right-moving electrons, which results in a negative current because of
the negative electron charge $e$. However, it should be noted that the
current can be positive for small values of $\hbar\omega$, as in the
$\varphi=3\pi /2$ curve in Fig.~\ref{fig2}(a) (Note that this curve
has been shifted vertically by $1.2 \cdot 10^{-11}$ A). This behavior
is due to quantum interference phenomena in the reflection and
transmission of the electrons across the three regions. As illustrated
in Fig. \ref{fig2}(a), the length of the irradiated region has a
significant effect on the photocurrent. In quantum wires with a
shorter arc, the current as a function of the photon energy is
characterized by a single peak at $\hbar \omega=\mu$. By increasing
the arc angle $\varphi$ additional peaks appears and the peak at
$\hbar \omega=\mu$ decreases. We found that the position of these
additional peaks is determined only by the arc length (at a fixed $R$)
and does not depend on the radiation intensity. Fig. \ref{fig2}(b)
demonstrates that the current at $\hbar \omega=\mu$ is stronger in the
wires with larger $R$ and its maximum shifts to smaller $\varphi$ with
increase of $R$.

The effects of the radiation intensity and finite temperature on the
photocurrent are shown in Fig.~\ref{fig3}. In Fig.~\ref{fig3}(a) we
see that the energy dependence of the photocurrent scales with the
radiation intensity without changing considerably in shape. The finite
temperature (Fig. \ref{fig3}(b)) smoothes the peak at $\hbar
\omega=\mu$ and shifts it to lower energy. However, this effect
becomes significant only at $T \sim 10$K, implying that very strict
temperature requirements are not needed in the experiment.

In summary, we have demonstrated that circularly polarized
electromagnetic radiation induces a current in curved ballistic
quantum wires (photovoltaic effect). The current was calculated as
a function of the photon energy and length of the bent segment.
We have investigated the temperature and intensity dependence of
the photocurrent.  We found that for a realistic set of parameters
a current of the order of $10$pA can be observed. In curved
quantum wires with a short curved segment, the current dependence
on photon energy shows a single peak at the Fermi energy. Larger
segments give rise to additional peaks due to the wave reflection
and transmission at different region boundaries.

We thank Prof. M. Dykman for many fruitful discussions. This
research was supported by the National Science Foundation, Grant
NSF DMR-0312491.

\end{document}